  \newcommand{\Hcal}{        {H}}

\def\slat{\mathcal{S}}
\def\teop{\mathcal{U}}
\def\dcoo{\mathcal{T}}
\def\MHam{\mathcal{H}}
  \def\Perfect{{\rm P}} 
  \def\Imperfect{{\rm L}} 
  \def\TQD{\rm B} 
  \def\mbqsp{\rm MB-QSP}
  \def\myVarPar{\Lambda}
  \def\myEEV{{\cal E}}
  \def\myEnDefFun{{\epsilon}}
  \def\myDub{{\mathbf u}}
  \def\myResponse{{\cal R}}
  \def\myHamMat{\mathbb{H}}
  \def\myUnit{{\cal I}}
  \def\myProj{{\cal P}}
  \def\myAstar{{A_{\rm opt.}}}
  \def\myLocTen{{\cal L}^{2}}
  
\def\ie{{\it i.e.\/}}

\def\viz{{\it viz.\/}}
\def\via{{\it via\/}}
\def\bra#1{\langle#1\vert}
\def\ket#1{\vert#1\rangle}

\def\me#1#2#3{\langle#1\vert#2\vert#3\rangle}
\def\myDet{{\rm Det}}
\def\myTr{{\rm Tr}}
\def\filledDet{\overline{\myDet}}
\def\filledTr{\overline{\myTr}}

\documentclass[aps,prl,twocolumn,groupedaddress]{revtex4-2}
\usepackage{amsmath, amsthm, amssymb, mathtools, thmtools,xcolor,dsfont}

\begin{document}


\title{Towards perfect quantum insulation}


\author{Rafael Hipolito$^{1,2}$}
\author{Paul M.~Goldbart$^{1}$}
\affiliation{$^{1}$Department of Physics and Astronomy, Stony Brook University}
\affiliation{$^{2}$Department of Physics, University of Texas at Austin}


\date{January 23, 2025}

\begin{abstract}
Electric fields, applied to insulators, cause transitions between valence and conduction bands, giving rise to current. Adjustments of the Hamiltonian can perfect the quality of the insulator, shutting down transitions whilst fully preserving the many-particle state, but they are challenging to implement. Instead, adjusted Hamiltonians having desirable features are addressed variationally, \via\ the analysis of a suitable figure of merit. They suppress current-enabling transitions whilst tending to preserve the many-particle state, and hence they yield optimal insulation. Emerging naturally from this approach are two established concepts: transitionless quantum driving 
[M.~V.~Berry, {\sl J.~Phys.~A: Math.~Theor\/}.~{\bf 42\/}, 365303 (2009)] 
and (a modified) localization tensor 
[R.~Resta and S.~Sorella, {\sl Phys.~Rev.~Lett.\/}~{\bf 82\/}, 370-373 (1999)].
The variational approach is illustrated \via\ application to a tight-binding model. In this setting, the optimally adjusted Hamiltonian has a powerful impact on transition suppression and localization-tensor reduction, suggesting strong enhancement of insulation. These features are expected to be more general than the model that displays them.
\end{abstract}


\maketitle

In the absence of quantum transitions, electrically insulating ground-states of matter respond to static external electric fields by developing electric polarization but not static electric current. Band insulators composed of non-interacting electrons provide the most familiar example of insulators but, as recognized following the work of Kohn~\cite{ref:WKohn1964} and others, there are other important scenarios, including Mott insulation, driven by electron-electron interactions, and Anderson insulation, driven by disorder. However, even at zero temperature, insulation is subject to breakdown~\cite{ref:LMZ}: the external electric field induces quantum transitions to current-carrying states, so that insulation is imperfect. In the band-insulation case, the current results from single-electron transitions from the (filled) valence band to the (empty) conduction band. 

In this Letter we aim to show that -- by invoking ideas rooted in Berry's concept of transitionless quantum driving~\cite{ref:berry2009} -- the transitions responsible for the breakdown of electrical insulation can be suppressed. In principle, this suppression can be complete, provided no limitations are placed on the complexity of the scheme used to achieve it. However, as we shall see, even under the constraint that the scheme be local, and therefore more readily implementable in practice, the rate at which transitions occur can be substantially diminished. 
Thus, the quality of the insulator, if not perfected, can be considerably improved {\it via\/} the addition of a certain local Hamiltonian that tends to {\it preserve the complete many-particle quantum state\/} 
by counteracting the tendency of the electric field to induce transitions to current-carrying states. 
Whilst our analysis focuses on the case of band insulators and non-interacting electrons, we expect its general features to be of broader relevance, including to Mott and Anderson insulators. 

Thus we consider spatially periodic systems of arbitrary dimensionality, populated by electrons whose inter-particle interactions we neglect. 
Although not essential, it is useful to envision a tight-binding description. 
We incorporate a spatially homogeneous but temporally arbitrary electric field \via\ suitably time-dependent hopping matrix elements. 
Regarding states, we allow for an arbitrary number of single-particle energy bands, separated by gaps. 
It is straightforward to include the spin of the electron, 
but because it does not play an essential role here we omit it.
As we are neglecting interactions, it is evident that at any instant $t$ there is a special set of many-particle states in which, for every band, either all of its co-moving single-particle states~\cite{fn:OnCoMo} are populated or all of them are empty. 
We call such states {\it inert\/} and denote them by 
$\ket{\slat(t)}$. They are Slater determinants built 
from the filled co-moving single-particle states; 
their dependence on time reflects the time-dependence 
of the (co-moving) single-particle states. It is these  
many-particle states $\ket{\slat(t)}$ that we are seeking 
to preserve. 
In the case of the one-dimensional, two-band model discussed in detail below, inert states occur at half filling: the (fermionic but spinless) electrons fill either the lower band (state $\ket{\slat_{1}(t)}$, \ie, the ground state) or the upper band (state $\ket{\slat_{2}(t)}$); in each case, the other band is empty. 

We now introduce a central entity: 
a {\it figure of merit\/} 
\begin{equation}
\Omega(t)\equiv
\big\vert
  \me{\slat(t)}{\,\teop(t)}{\,\slat(0)}\big\vert^{2}
\label{eq:FOMdef}
\end{equation}
for any inert state $\ket{\slat(t)}$, with which we quantify the extent to which the complete many-particle quantum state 
$\ket{\slat(0)}$ is preserved after a time $t$ 
under a generic unitary time-evolution $\teop$: 
$\Omega(t)=1$ would indicate perfect preservation; 
the amount by which $\Omega(t)$ falls below 1 
would quantify how impaired preservation is. 
For the present purposes, three distinct time-evolutions are of interest. 
\hfil\break\noindent 
(i)~Evolution generated by the time-dependent many-body Hamiltonian $\MHam_{0}(t)$, 
which incorporates the effects of the lattice and the electric field. 
We denote this evolution by $\teop_{0}(t)$; it is given by 
$\dcoo\exp\big[ -(i/\hbar)
\int_{0}^{t}d\tau\,\MHam_{0}(\tau)\big]$, 
where $\dcoo$ is the standard time-ordering operation.  
The time dependence of $\MHam_{0}(t)$ brings the transitions 
that tend to deplete the initially filled band(s) 
[so that $\Omega(t)$ becomes less than unity]. 
At this stage, no effects have been introduced to allay depletion. 
\hfil\break\noindent
(ii)~Many-body quantum-state-preserving (\mbqsp) evolution, which we denote by $\teop_{0+\Perfect}(t)$. 
This evolution can be generated by any member of a family of {\it adjusted\/} many-particle Hamiltonians, $\MHam_{0}(t)+\MHam_{\Perfect}(t)$, in which the family of possible terms $\{\MHam_{\Perfect}(t)\}$ is determined by $\MHam_{0}(t)$ and the state to be preserved, as we explain below. 
One member of this family, $\MHam_{\TQD}(t)$, is the term that would ensure transitionless quantum driving in precisely the form delineated by Berry~\cite{ref:berry2009}. 
The form of state preservation that we are seeking to accomplish has greater freedom than Berry's. 
This is because it does not require that each {\it individual\/} co-moving single-particle state be transitionless. 
Rather, we only seek to preserve some specified (in our case, inert) many-particle quantum state. 
Any member of the aforementioned family exactly ensures such preservation by supplying only the necessary suppression of transitions, \ie, the transitions between filled and unfilled states (and not every transitions that the time-dependence of $\MHam_{0}(t)$ would alone induce).  
With $\MHam_{\Perfect}(t)$ present, 
then, $\Omega(t)=1$: the complete many-body state -- and not just its occupation numbers -- would be perfectly preserved. 
As we further discuss below, even if $\MHam_{0}$ were local 
there would be no reason to expect $\MHam_{\Perfect}$ to be local~\cite{fn:locality}. 
Thus, in practice it may well be difficult to achieve perfect preservation of the many-particle state, and for this reason we consider a third evolution. 
\hfil\break\noindent
(iii)~Here, we give up on perfect preservation, asking instead: 
How well can we preserve the many-particle state if we exchange 
$\MHam_{\Perfect}$ for a parametrized class of plausibly achievable 
(which in the illustrative example we explore below means local) 
Hamiltonians $\MHam_{\Imperfect}[\myVarPar(\cdot)]\,$, 
and optimize $\Omega$ with respect to the time-dependent parameters $\myVarPar(\cdot)$? 
This would give not \mbqsp\ but its best approximant, so that 
$\Omega[\MHam_{\Imperfect}]$ would obey 
$\Omega[\MHam_{\Imperfect}=0]<\Omega[\MHam_{\Imperfect}]<\Omega[\MHam_{\Perfect}]$ 
with $\Omega[\MHam_{\Imperfect}]$ as large as possible. 

Focusing, then, on evolution~(iii), and restricting our attention to the case of non-interacting electrons, we now explain how the figure of merit can be expressed in the useful form 
(writing $\Omega_{\Imperfect}$ for $\Omega[\MHam_{\Imperfect}]$): 
\begin{equation}
\Omega_{\Imperfect}(t)=
\big\vert
{\filledDet}\,\,
\overline{\teop}_{\Imperfect-\Perfect}(t)
\big\vert^{2}\!=
{\filledDet}\,\big(\,
\overline{\teop}_{\Imperfect-\Perfect}^{\dagger}(t)\,\,
\overline{\teop}_{\Imperfect-\Perfect}(t)\big).
\label{eq:diffham}
\end{equation}
Here, $\overline{\teop}_{\Imperfect-\Perfect}(t)$ represents time evolution $\teop_{\Imperfect-\Perfect}$ 
according to $\MHam_{\Imperfect}-\MHam_{\Perfect}$, but with $\teop_{\Imperfect-\Perfect}$ projected on to the 
reduced space spanned by the filled single-particle states in $\ket{\slat}$. In addition, $\filledDet$ indicates a 
determinant, also taken in the reduced space, according to the point made in Ref.~\cite{fn:WhatBasisQ}.
Note that $\overline{\teop}_{\Imperfect-\Perfect}$ is not unitary, 
except in the trivial case in which all single-particle states are filled~\cite{fn:IfTQD}. 
One can obtain Eq.~(\ref{eq:diffham}) in the following way. 
Start with the matrix element $\me{\slat(t)}{\,\teop(t)}{\,\slat(0)}$ of Eq.~(\ref{eq:FOMdef}). 
Develop a path-integral representation by introducing fermion coherent-state resolutions of the 
(Fock-space) identity at each intermediate time slice. 
Unusually, however, choose the single-particle 
basis at each time slice not to be a fixed basis but, rather, the co-moving basis corresponding to that 
particular time. This is where the physics of adiabaticity comes in. 
At each time slice we manage the variation in basis \via\ a suitable unitary transition function 
(\ie, connection). As a consequence, the Hamiltonian 
$\MHam_{0}+\MHam_{\Imperfect}$ 
acquires the additional contribution 
$-\MHam_{0}-\MHam_{\Perfect}$, 
thus becoming
$\MHam_{\Imperfect}-\MHam_{\Perfect}\,$~\cite{fn:DPInclude}. 
One special choice of transition function (\ie, the one that implements 
Berry's state-by-state transitionless driving~\cite{ref:berry2009}) gives rise to 
\begin{equation}
\MHam_{\Perfect}=i\hbar\sum\nolimits_{n}\left(\partial_{t}\ket{n(t)}\right)\bra{n(t)}\,,
\label{eq:HamforTQD}
\end{equation}
where $\{\ket{n(t)}\}$ are the co-moving eigenstates of the time-dependent Hamiltonian $\MHam_{0}$. 
In the present setting, this particular $\MHam_{\Perfect}$ is but one out of the entire family $\{\MHam_{\Perfect}\}$ 
generated by the choice of co-moving basis at each time-slice. 
This flexibility (in essence a gauge freedom) arises because {\it all\/} mixings that do not mix filled and empty states in $\ket{\slat}$ are allowed.  
Mixing within the filled (or empty) states leaves them filled (or empty), and therefore has no physical effect. 
It is attractive that $\{\MHam_{\Perfect}\}$ emerges without having been introduced by hand. 
Moreover, the combination $\MHam_{\Imperfect}-\MHam_{\Perfect}$ itself is satisfying, 
inasmuch as it is departures of $\MHam_{\Imperfect}$ 
from $\MHam_{\Perfect}$ that give rise to transitions that 
would deplete filled single-particle states. 
The particular realization of $\MHam_{\Perfect}$ that arises (from the family $\{\MHam_{\Perfect}\}$) depends on the particular choice of how the co-moving basis evolves. However, the physics is independent of gauge choice.

We introduced the figure of merit $\Omega(t)$ to quantify state preservation at time $t$. 
We now harness this property in a variational sense mentioned above to identify 
the optimally state-preserving Hamiltonian $\MHam_{\Imperfect}\,$. 
Correspondingly, Eq.~(\ref{eq:diffham}) gives rise to a figure of merit 
$\Omega_{\Imperfect}(t)$ that depends functionally on $\myVarPar$, 
and we seek optimality using stationarity, 
${\delta}\Omega_{\Imperfect}=0$, 
with respect to variations ${\delta}\myVarPar$. 
Thus, we arrive at the condition: 
\begin{equation}
\delta{\Omega}_{\Imperfect}
={\Omega}_{\Imperfect}\,
\filledTr\,\big[\!\left(\,\overline{\teop}_{\Imperfect-\Perfect}\right)^{-1}
\delta\,\overline{\teop}_{\Imperfect-\Perfect}+{\rm h.c.}\big]=0\,, 
\end{equation}
again noting the point made in Ref.~\cite{fn:WhatBasisQ}.
Here, all terms are taken at the same time $t$, 
$\overline{\teop}_{\Imperfect-\Perfect}$ is the projection (on to the space spanned by the filled states) 
of the evolution driven by $\MHam_{\Imperfect}-\MHam_{\Perfect}\,$, 
and $\delta\,\overline{\teop}_{\Imperfect-\Perfect}$ is the variation in $\overline{\teop}_{\Imperfect-\Perfect}$
arising from the ($\myVarPar$-induced) variation in $\MHam_{\Imperfect}\,$.

With the formalism of the figure of merit now developed, 
we turn to our primary aim, which is to determine the optimal $\myVarPar$ and its consequences.
With this aim in mind, it is useful to recast Eq.~(\ref{eq:diffham}) 
in a way that highlights the role played by transitions between 
filled and unfilled single-particle states. 
This brings us to the equivalent formula: 
\begin{equation}
{\Omega}_{\Imperfect}(t)\!=\!
\exp\!\myTr\ln\!\big[
\myUnit\!-\!
\big(\myUnit\!-\!\overline{\myProj}(t)\big)
\,{\teop}_{0+\Imperfect}(t)
\,\overline{\myProj}(0)
\,{\teop}_{0+\Imperfect}^{\dagger}(t)
\big], 
\label{eq:ExpForm}
\end{equation}
where, for the sake of clarity, we emphasize that both the trace and the evolution operators 
refer to the full space, not the reduced space, 
and $\overline{\myProj}(t)$ denotes the projection operator on to the inert state $\ket{\slat}$ at time $t$. Moreover, the trace is now an ordinary one; 
the projection operators take care of the co-movement of the single-particle states. 
Equation~(\ref{eq:ExpForm}) follows from Eq.~(\ref{eq:diffham}) by 
using the standard relation $\ln\filledDet=\filledTr\ln$ 
and then expressing $\overline{\teop}_{\Imperfect-\Perfect}(t)$ in terms of 
${\teop}_{\Imperfect-\Perfect}(t)$ and projection operators. 
This brings in the full-space evolution operator, which is advantageous, 
owing to its direct connection to the operative Hamiltonian. 
The presence of the projection operators is also advantageous: it makes explicit the 
essential role played by transitions between filled and unfilled single-particle 
states and the irrelevance of all other transitions. 

To acquire an understanding of the effectiveness of our transition-mitigating scheme, 
it is useful to consider the early-stage behavior of ${\Omega}_{\Imperfect}(t)$. 
Indeed, by examining Eq.~(\ref{eq:ExpForm}) 
in the light of the paragraph that follows it, 
one finds that the term 
$\big(\myUnit\!-\!\overline{\myProj}(t)\big)
\,{\teop}_{0+\Imperfect}(t)\,\overline{\myProj}(0)
\,{\teop}_{0+\Imperfect}^{\dagger}(t)$ 
grows quadratically in $t$, 
so that for sufficiently small times 
we may expand the logarithm, thus obtaining 
\begin{equation}
{\Omega}_{\Imperfect}(t)\approx
\exp\!\big[-{\textstyle\left(\frac{t}{\hbar}\right)^{2}}\,\myTr\,
(\MHam_{\Imperfect}\!-\!\MHam_{\Perfect})
(\myUnit\!-\!\overline{\myProj})
(\MHam_{\Imperfect}\!-\!\MHam_{\Perfect})
\overline{\myProj}\,\big],
\label{eq:ExpEpandForm}
\end{equation}
in which all operators are taken at $t=0$. 
(Time-reversal invariance of ${\Omega}_{\Imperfect}$ ensures 
that no term linear in $t$ arises in the exponent.)\thinspace\ 
Let us pause to consider the physical significance of the exponent factor
\begin{equation}
\myTr\,(\MHam_{\Imperfect}\!-\!\MHam_{\Perfect})
(\myUnit\!-\!\overline{\myProj})
(\MHam_{\Imperfect}\!-\!\MHam_{\Perfect})
\overline{\myProj}, 
\label{eq:Response}
\end{equation} 
which we term the {\it breakdown metric\/} 
and for future convenience write as: 
$({\dot{\phi}}\,\hbar)^{2}\,M\,\myResponse_{\Imperfect}\,$.
We already know that 
${\Omega}_{\Imperfect}(t)=1$ 
whenever $\MHam_{\Imperfect}$ is drawn from the family 
$\{\MHam_{\Perfect}\}$ and that 
${\Omega}_{\Imperfect}(t)<1$ 
for all other choices of $\MHam_{\Imperfect}$ (including zero), 
as discussed between Eqs.~(\ref{eq:FOMdef}) and (\ref{eq:diffham}).
Thus, we recognize the (reduced) breakdown metric 
$\myResponse_{\Imperfect}$
as a diagnostic of the extent to which time-evolution fails to preserve some specified inert many-particle state. 
Inasmuch as it would 
vanish for a perfect insulator, 
grow as the quality of the insulator is reduced 
(and, we presume, ultimately diverge as the transition to a metal is approached), 
it bears the hallmarks of the localization tensor 
introduced by Resta and Sorella~\cite{ref:RandSprl1999} 
as a diagnostic of the insulating state 
and also featuring centrally in the work of Marzari and Vanderbilt 
on maximally localized Wannier functions~\cite{ref:MazVan1997}; 
for reviews, see Refs.~\cite{ref:RR-reviews}.
In fact, as we shall see below, 
when $\MHam_{\Imperfect}=0$ the intensive form of the exponent 
$\myResponse_{\Imperfect}[\MHam_{\Imperfect}=0]$
{\it is\/} the localization tensor. 
Furthermore, it continues to diagnose the quality of the insulator even when $\MHam_{\Imperfect}\ne 0$ as it then captures the improvement in insulation as $\MHam_{\Imperfect}$ is optimized.

To illustrate these ideas, we examine in detail the case of a tight-binding model on a 
one-dimensional lattice that has two sites per unit cell. (The program we implement is 
readily generalizable to systems having larger numbers of dimensions and/or orbitals 
per unit cell.)\thinspace\ Thus, we consider the single-particle Hamiltonian 
\begin{equation}
\Hcal_0(t)\equiv
\sum_{m=0}^{M-1}[-J{\rm e}^{i\phi(t)}|m\rangle\langle m+1| +{\rm h.c.\/}] 
+(-1)^m\Delta |m\rangle \langle m|,
\label{eq:lattice}
\end{equation}
where $m$ ($=0,1,\ldots,M-1$) labels the sites of the lattice, 
the (even) number $M$ is the total number of sites, 
$J$ is a (real-valued) hopping matrix element, 
the dependence on time $t$ of the phase factor $\exp{i\phi(t)}$ 
captures the effect of a spatially uniform applied electric field, 
and $2\Delta$ specifies the difference between the on-site energies 
of the even-$m$ and odd-$m$ sites.  
With its pair of sites per unit cell, at any instant $t$ the solution of the 
single-particle energy eigenproblem for the periodic Hamiltonian $\Hcal_0(t)$ 
consists of two bands, labeled $-$ (the lower) and $+$ (the upper), 
comprising $M/2$ states and 
separated by a band gap $2\Delta$.  
The corresponding energy eigenstates and eigenvalues are given by
\begin{subequations}
\begin{eqnarray}
\ket{\psi_{\pm,\kappa}}&=&
\sqrt{\frac{2}{M}}\sum_{m=0}^{M-1}
{\rm e}^{i\kappa m} u_{\pm,k(t)}(m)\ket{m},
\\
\begin{bmatrix}
u_{\pm,k}(0)\\ 
u_{\pm,k}(1)
\end{bmatrix}
&=& 
\frac{1}{\sqrt{2\epsilon_{k}\,[\myEnDefFun_{k}\pm\Delta]}}
\begin{bmatrix}
\Delta\pm\myEnDefFun_{k}
\\
-2J\cos{k}
\end{bmatrix},
\end{eqnarray}%
\label{eq:Spectrum}%
\end{subequations}%
where $\myEEV_{\pm,k}=\pm\myEnDefFun_{k}$
and $\myEnDefFun_{k}\equiv\sqrt{\Delta^{2}+4J^{2}\cos^{2}{k}}$. 
In addition, 
$\kappa$ [$=2\pi\mu/M$, with $\mu=0,1,\ldots,(M/2)-1$] 
is the quasi-momentum, 
$k(t)$ [$\equiv\kappa+\phi(t)$] 
is the field-shifted quasi-momentum, 
and $u_{\pm,k(t)}(m)$ [$=u_{\pm,k(t)}(m+2)$]
are the Bloch wave functions~\cite{fn:PerBC}. 
Note that we have chosen units in which the inter-site distance is unity. 
As time proceeds, the states $\ket{\psi_{\pm,\kappa}}$ flow parametrically through the Brillouin zone~\cite{fn:gaugechoice}; 
the sets of eigenfunctions and eigenvalues recur each time $\phi$ changes by a multiple of $2\pi/M$. 
It is useful to combine 
$u_{\pm,k}(0)$ and 
$u_{\pm,k}(1)$ 
into the doublet 
$\myDub_{\pm,k}\equiv[u_{\pm,k}(0),u_{\pm,k}(1)]^{\rm T}$, 
with normalization 
$\myDub_{\pm,k}^{\dagger}\cdot\myDub_{\pm,k}^{\phantom\dagger}=1$. 
We focus on the particular inert state in which the lower band is filled with  
electrons and the upper band is empty. The electric field, entering \via\ the 
time-dependence of $\phi(t)$, tends to induce electron transitions into the upper band. 

We now consider a variational family of Hamiltonians $\MHam_{\Imperfect}[\myVarPar]\,$, 
guided by the following requirements, which are particular to this illustrative example. 
(i)~We continue to exclude electron-electron interactions, 
in which case it is adequate to specify a variational single-particle Hamiltonian 
${\Hcal}_{\Imperfect}(t)$. 
(ii)~We maintain the translational invariance of the lattice. 
(iii)~We admit on-site terms and nearest-neighbor hopping terms, 
respectively characterized by the variational parameters $V_{m}(t)$ and $A_{m}(t)$, 
consistent with the notion that $\Hcal_{\Imperfect}(t)$ be plausibly achievable. 
Thus, we are led to the term: 
\begin{eqnarray}
\Hcal_{\Imperfect}
&=& 
\hbar\,\dot{\phi}
\sum_{m=0}^{M-1}
\Big[V_{m}\ket{m}\bra{m}+
\big(A_{m}\ket{m+1}\bra{m}+{\rm h.c.}\big)\Big],
\nonumber
\\
{\rm where}&&
V_{m+2} = V_{m}
\quad{\rm and}\quad
A_{m+2}=A_{m}\,,
\label{eq:BLAparam}
\end{eqnarray}
where it is convenient to extract the factor $\dot{\phi}$. 
The first term in $\Hcal_{\Imperfect}$ 
amounts to a renormalization of the staggered on-site potential; 
the second to a staggered renormalization of the hopping.  
What remains is to determine the optimal values of the two 
real parameters $(V_0,V_1)$ and 
complex parameters $(A_0,A_1)$.  

Using Eq.~(\ref{eq:ExpEpandForm}), 
we now address the short-time behavior 
of $\Omega_{\Imperfect}$ for the 
model~(\ref{eq:lattice}), 
variational Hamiltonian~(\ref{eq:BLAparam}), 
and choice of gauge (\ie, transition function) 
that yields the particular form for $\{\MHam_{\Perfect}\}$ given in Eq.~(\ref{eq:HamforTQD}). 
Using the eigenvalues and eigenfunctions 
given in Eqs.~(\ref{eq:Spectrum}), we obtain, in the large-system limit, 
\begin{subequations}
\begin{eqnarray}
\Omega_{\Imperfect}
&\approx&
\exp\,[-({\dot{\phi}}\,t)^{2}\,M\,\myResponse_{\Imperfect}],
\label{eq:OmExp}
\\
\noalign{\smallskip}
\myResponse_{\Imperfect}
\!&\equiv&\!
\int_{-\pi/2}^{\pi/2}\frac{dk}{2\pi}\,
\big\vert
\myDub_{+,k}^{\dagger}
\cdot\!\big[
\myHamMat_{\Imperfect,k}-i\partial_{k}
\big]\!\cdot
\myDub_{-,k}^{\phantom\dagger}
\big\vert^{2},
\label{eq:respo}
\\
\noalign{\smallskip}
\myHamMat_{\Imperfect,k}
&\equiv&
\begin{bmatrix}
V_0 & {\rm e}^{ik}A_0^{*}+{\rm e}^{-ik} A_1 
\\
{\rm e}^{-ik}A_0 + {\rm e}^{ik}A_1^{*} & V_1
\end{bmatrix}. 
\label{eq:Hmat}
\end{eqnarray}
\end{subequations}
In this limit, we can make the standard replacement: 
$\partial/\partial\phi\rightarrow\partial/\partial k$, 
which is applicable to $\mathbf{u}_{\pm,k}$ since it only depends on the combination $k(t)=\kappa+\phi(t)$.
It is straightforward to carry out the resulting 
wave-vector integrals in Eq.~(\ref{eq:respo}).  
In doing so, it becomes evident that  
minimizing $\myResponse_{\Imperfect}$ 
(and hence maximizing $\Omega_{\Imperfect}$) 
requires 
$(V_0,V_1)=(0,0)$ 
(up to a physically irrelevant common shift) 
and 
$(A_0,A_1)=(-1,1)A$ (with $A$ real). 
The result for $\myResponse_{\Imperfect}$
depends on the band-width $2J$ and band-gap $\Delta$ 
only \via\ their ratio $\xi$ $[\equiv 2J/\Delta]$: 
\begin{eqnarray}
\myResponse_{\Imperfect}
&=& 
\frac{\xi^{2}}{16\sqrt{1+\xi^{2}}} - 
\frac{A}{\xi}
\big[\sqrt{1+\xi^{2}}-1\big] 
+ A^{2}. 
\end{eqnarray}
Hence, we find that the value of $A$  
that best suppresses transitions 
and thus best preserves the quantum many-particle state 
(\viz, $\myAstar$)
is given by 
\begin{eqnarray}
\myAstar=
\frac{1}{2\xi}\big[\sqrt{1+\xi^2}-1\big],
\end{eqnarray}
and therefore that the optimal value of $\myResponse_{\Imperfect}$ 
takes the value 
\begin{equation}
\myResponse_{\rm opt.}=
\frac{\xi^{2}}{16\sqrt{1+\xi^{2}}} - 
\frac{1}{4\xi^{2}}\,
\big[\sqrt{1+\xi^2}-1\big]^{2}.
\label{eq:OptRespo}
\end{equation} 
Recall that our aim is to understand how and how well an inert many-particle quantum state can be preserved 
(in the face of time-dependence that induces quantum transitions)
\via\ the application of 
an approximation, $\MHam_{\Imperfect}$, to 
the family of ideal Hamiltonians $\{\MHam_{\Perfect}\}$. 
We can gain a sense of the efficacy of the preservation strategy by continuing with the one-dimensional, two-band model and considering the limit in which the band gap is much larger than the band width, \ie, $\xi\ll 1$. This is the limit in which transitions are weak, even when unmitigated by any preservation strategy. (Said more colloquially, the system is already a good insulator.)\thinspace\ 
In this regime, to leading order in ${J}/{\Delta}$ we find that 
\begin{equation}
\myAstar={\textstyle\frac{1}{2}}({J}/{\Delta})
+{\cal O}({J}/{\Delta})^{3},
\end{equation}
so that 
(i)~absent any preservation strategy or 
(ii)~with the optimal version of it we find, respectively, that 
\begin{equation}
\!\!\!\!
\myResponse_{\Imperfect}
\approx
\begin{cases}
{\rm \phantom{i}i.}\,\,\,
{\frac{1}{4}\left({J}/{\Delta}\right)^{2}}
+{\cal O}({J}/{\Delta})^{4}, 
\,\,\,
\textrm{for~$A=0\,$}; 
\\
\noalign{\smallskip}
{\rm ii.}\,\,\,
\frac{1}{4}\left({J}/{\Delta}\right)^{6}
+{\cal O}({J}/{\Delta})^{8}, 
\,\,\,
\textrm{for~$A=\myAstar$}.
\end{cases}
\end{equation}
Thus, for this particular operative regime 
(\viz, $J\ll\Delta$), 
one finds a striking improvement of the state-preservation efficacy, as signaled by the increase in the exponent governing the $(J/\Delta)$-dependence, from $2$ to $6$, and a correspondingly weaker time-decay behavior of the figure of merit $\Omega$. 

We now return to the breakdown metric defined in 
Eq.~(\ref{eq:Response}). 
Evaluating it in the setting of any one-dimen{\-}sional periodic system of noninteracting electrons in any inert state in which only a single band (the $n^{\rm th}$) is filled, one finds
\begin{eqnarray}
\!\!\!\!\!\!\!\!
&&
\myResponse_{\Imperfect}[\MHam_{\Imperfect}=0]=
{\frac{1}{M\hbar^{2}}}\,\myLocTen,
\\
\noalign{\smallskip}
\!\!\!\!\!\!\!\!
&&
\myLocTen=
\!\!\!
\sum_{k\,{\rm in\,BZ}}\!\!\!
\big[
\bra{\psi_{n,k}}
\partial_{k}^{2}
\ket{\psi_{n,k}}
-\!\!\!
\sum_{q\,{\rm in\,BZ}}\!\!
\big\vert
\bra{\psi_{n,q}}
\partial_{k}
\ket{\psi_{n,k}}
\big\vert^{2}
\,\big], 
\label{eq:LocTenSums}
\end{eqnarray}
where $\{\ket{\psi_{n,k}}\}$ are single-particle Bloch states of wave-vector $k$~\cite{fn:MoreBands}.
(The extension to higher dimensions is straightforward.)\thinspace\ 
As anticipated below Eq.~(\ref{eq:Response}), what emerges is the Resta-Sorella localization tensor. 
Application of the optimal $\MHam_{\Imperfect}$ yields the minimum possible breakdown metric, in effect, optimally reducing the localization tensor.

We have addressed insulating states and strategies for preserving them in the presence of electric fields. 
Whilst in principle this can be done perfectly, in practice this requires the application of Hamiltonians that are likely to be difficult to realize. 
With this in mind, we have developed an optimization strategy which focuses on a convenient figure of merit that indicates how well state-preservation is attained when the system is subjected to an additional, adjusting Hamiltonian chosen optimally from a family that is intended to be feasible. 
Along the way, Berry's transitionless driving prescription arises naturally as one exemplar of a broader class of Hamiltonians that perfectly preserve insulating states. 
Naturally arising, too, is the Resta-Sorella localization tensor, in connection with the susceptibility to electric-field-induced transitions that the figure of merit quantifies. 
The figure of merit also reveals how the localization tensor changes (and indeed shrinks) in the presence of optimal adjusting Hamiltonians that are selected to suppress transitions and thus promote insulation. 

The formalism presented here is not restricted to charge transport in electronic systems, but may be applied to any setting in which many-freedom quantum-state--preservation -- and how well it can feasibly be realized -- is the goal. Whatever the setting, the formalism would yield the appropriate analog of the localization tensor as well as how the appropriate breakdown metric would change as a reflection of the extent of transition suppression. 
Although we have exhibited the formalism in the linear-response (or short-time) regime, we note that it is not thus restricted and may therefore be applied more broadly. 

Not discussed here but, we believe, worth addressing in the future are applications to the Schwinger strong-field pair-creation mechanism~\cite{ref:JSchwinger-1951} and to the precision of quantization of particle delivery by quantum pumps operating at a non-zero rate. Also worth addressing are interacting systems, including Mott insulators, and disordered systems, such as Anderson insulators. 

\begin{acknowledgments}
This work was performed in part at the Aspen Center for Physics, which is supported by National Science Foundation grant PHY-2210452.
\end{acknowledgments}

{\raggedright 
}


\begin{thebibliography}{99}

\bibitem{ref:WKohn1964}
W.\ Kohn, 
{\sl Phys.\ Rev.\/}~{\bf 133\/}, A171-A181 (1964).
\bibitem{ref:LMZ}
See Refs.~\cite{ref:landau1932,ref:majorana1932,ref:zener1932}. 
\bibitem{ref:landau1932}
L.\ Landau, 
{\sl Phys.\ Z.\ Sow.\/}~{\bf 2\/}, 46-51 (1932).
\bibitem{ref:majorana1932}
E.~Majorana, 
{\sl Nuovo Cim.\/}~{\bf 9\/}, 43-50 (1932).
\bibitem{ref:zener1932}
C.\ Zener, 
{\sl Proc.\ R.\ Soc.\ Lond.\/}~{\bf A~137\/}, 696-702 (1932).  
\bibitem{ref:berry2009}
M.\ V.\ Berry, 
{\sl J.\ Phys.\ A: Math.\ Theor\/}.~{\bf 42\/}, 365303 (2009) [9~pages]. 
\bibitem{fn:OnCoMo}
By co-moving states we mean instantaneous eigenstates of the Hamiltonian for the spatially periodic system and spatially homogeneous but time-dependent electric field. For convenience, we include the associated dynamical phases in the definition of the co-moving states. 
\bibitem{ref:Wannier1937}
G.\ H.\ Wannier, 
{\sl Phys.\ Rev.\/}~{\bf 52\/}, 191-197 (1937). 
\bibitem{ref:Wannier1960}
G.\ H.\ Wannier, 
{\sl Phys.\ Rev.\/}~{\bf 117\/}, 432-439 (1960).
\bibitem{fn:locality}
In the present context, by {\it local\/} we mean a single-particle 
operator that involves only on-site and nearest-neighbor terms. 
We single out this class of terms because it seems feasible 
to realize them, at least in contrast with the fully mitigating 
operator $\MHam_{\Perfect}$.  
\bibitem{fn:WhatBasisQ}
Unusually, the matrix whose determinant or trace is 
being taken (whether in the full or reduced space) consists 
of the matrix elements of the appropriate operator, not taken 
between a single-particle basis of bras and the dual basis of 
kets but, rather, between co-moving versions of these, at 
time $t$ for the bras and time $0$ for the kets. 
\bibitem{fn:IfTQD}
$\overline{\teop}_{\Imperfect-\Perfect}$ 
would also be unitary if 
$\MHam_{\Imperfect}-\MHam_{\Perfect}$ 
were a member of the class of Hamiltonians  
$\{\MHam_{\Perfect}\}$. 
\bibitem{fn:DPInclude}
The contribution $-\MHam_{0}$ results from our choice 
of including dynamical phases in the definitions of the 
co-moving basis states.  
\bibitem{fn:gaugechoice}
For a choice of gauge corresponding to Eq.~(\ref{eq:HamforTQD}).
\bibitem{ref:RandSprl1999}
R.\ Resta and S.\ Sorella, 
{\sl Phys.\ Rev.\ Lett.\/}~{\bf 82\/}, 370-373 (1999).
\bibitem{ref:MazVan1997}
N.\ Marzari and D.\ Vanderbilt, 
{\sl Phys.\ Rev.\/}~B~{\bf 56\/}, 12847-12865 (1997).
\bibitem{ref:RR-reviews}
R.\ Resta, 
{\sl J.\ Phys.: Condens.\ Matter\/}~{\bf 14\/} R625–R656,  (2002); 
{\sl J.\ Phys.: Condens.\ Matter\/}~{\bf 22\/}, 123201 [19~pages] (2010); 
{\sl Riv.\ Nuovo Cimento\/}~{\bf 41\/}, 463-512 (2018).
\bibitem{fn:PerBC} 
Periodic boundary conditions identify the lattice-site 
pair $m=0$ and $m=1$ with the 
pair $m=M$ and $m=M+1$, 
respectively.
\bibitem{fn:MoreBands}
To obtain the localization tensor for the case of an inert state in 
which more than one band is filled, the summations over the Brillouin 
zone in Eq.~(\ref{eq:LocTenSums}) should be repeated for each filled band. 
\bibitem{ref:JSchwinger-1951}
J.\ Schwinger, 
{\sl Phys.\ Rev.\/}~{\bf 82\/}, 664-679 (1951).
\end{thebibliography}
\end{document}